\documentstyle[12pt,epsf]{article}\textheight 230mm\textwidth 160mm
\newcommand{\bi}{\bibitem}
\newcommand{\be}{\begin{eqnarray}}
\newcommand{\ee}{\end{eqnarray}}
\newcommand{\nn}{\nonumber}
\catcode`\@=11
\def\lsim{\mathrel{\mathpalette\@versim<}}
\def\gsim{\mathrel{\mathpalette\@versim>}}
\def\@versim#1#2{\vcenter{\offinterlineskip
\ialign{$\m@th#1\hfil##\hfil$\crcr#2\crcr\sim\crcr } }}
\catcode`\@=12
\begin{document}
\hspace*{9.7cm}\vspace{-3mm}KANAZAWA-95-18

\hspace*{9.7cm}October 1995
\begin{center}
{\Large\bf  The Top-Bottom  Hierarchy \\
from \\ Gauge-Yukawa Unification$\ ^{\dag}$}
\end{center}

\vspace{1cm}

\begin{center}{\sc Jisuke Kubo}$\ ^{1}$,
{\sc Myriam Mondrag{\' o}n}$\ ^{2}$,\\
{\sc George Zoupanos}$\ ^{3,**}$
\end{center}
\begin{center}
{\em $\ ^{(1)}$
 College of Liberal Arts, Kanazawa \vspace{-2mm} University,
Kanazawa 920-11, Japan } \\
{\em $\ ^{(2)}$ Institut f{\" u}r Theoretische Physik,
Philosophenweg 16 \vspace{-2mm}\\
D-69120 Heidelberg, Germany}\vspace{-2mm}\\
{\em $\ ^{(3)}$ Physics Department, National
Technical\vspace{-2mm} University\\ GR-157 80 Zografou, Athens,
Greece,  }
\end{center}

\vspace{1cm}
\noindent
{\sc\large Abstract}
\newline
\noindent
The idea of Gauge-Yukawa Unification (GYU)
based on the principle of reduction of couplings is
elucidated. We show how
 the observed top-bottom mass hierarchy
can be  explained
in terms of supersymmetric GYU by considering
an example of the minimal supersymmetric GUT.

\vspace*{4cm}
\footnoterule
\vspace*{2mm}
\noindent
$^{\dag}$ Presented by J.K. at the Yukawa International Seminar '95,
August 21-25, 1995, Kyoto, to appear in the proceedings.

\noindent
$^{**}$
Partially supported by  C.E.C. projects, SC1-CT91-0729;
CHRX-CT93-0319.
\newpage
\pagestyle{plain}

\section{Introduction}

Gauge-Yukawa Unification (GYU) is a functional
relationship among the gauge and Yukawa couplings, which
can be derived from some principle. Recall
 that the gauge and Yukawa
sectors in  Grand Unified Theories GUTs \cite{gut1} are usually not
related. But, in superstring and composite models for instance,
such relations
could be derived in principle.
In the GYU scheme
\cite{kubo1,mondragon1,kubo2}, which is based on the principle
of finiteness \cite{finite1,finite2} and reduction
of couplings \cite{zimmermann1},  one can write
down relations among the gauge and Yukawa couplings
 in a concrete fashion.
These principles are formulated
within the framework of
perturbatively renormalizable field theory, and
one can reduce the number of  independent
couplings without introducing necessarily a symmetry,
thereby improving the
calculability and predictive power of
a given theory \cite{zimmermann1}.

The consequence of GYU is that
in the lowest order in perturbation theory
 the gauge and Yukawa couplings
above the unification scale  $M_{\rm GUT}$
are related  in the form
\be
g_i& = &\kappa_i \,g_{\rm GUT}~,~i=1,2,3,e,\cdots,\tau,b,t~,
\ee
where $g_i~(i=1,\cdots,t)$ stand for the gauge
and Yukawa couplings, $g_{\rm GUT}$ for the unified coupling,
and
we have neglected  the Cabibbo-Kobayashi-Maskawa mixing
of the quarks.
Eq. (1) exhibits a boundary condition on the
renormalization group (RG) evolution for the effective theory
below $M_{\rm GUT}$, which we assume to
be the minimal supersymmetric standard model (MSSM).
 It has been recently found
\cite{mondragon1,kubo2} that various
supersymmetric GUTs with GYU in the
third generation can predict the bottom and top
quark masses that are consistent with the experimental data.
This means that the top-bottom hierarchy
could be
explained in these models,
exactly in the same way as
the hierarchy of the gauge couplings of the
standard model (SM)
can be explained if one assumes  the existence of a unifying
gauge symmetry above $M_{\rm GUT}$  \cite{gut2}.

Here we would like to outline the general idea of GYU which is based
on the principle of reduction of
couplings, and consider
a concrete example \cite{kubo2} to illustrate it.
Then we will briefly mention the idea of Dynamical Unification
of Couplings (DUC) that has been recently
proposed by one of us (J.K.) \cite{kubo3}
to understand a possible, theoretical origin of
reduction of couplings.

\section{GYU based on the principle of reduction of couplings}

Suppose  we have a set of couplings $\{ g_0,\cdots,g_N \}$.
(It is often
convenient to work with
$\alpha_{i}~\equiv~|g_i|^2/4\pi$.)
The principle of
reduction of couplings is to impose
 as many as possible RG invariant constraints
which are compatible with renormalizability \cite{zimmermann1}.
Such constraints in the space of couplings can be expressed in the
implicit form  as $\Phi (\alpha_0,\cdots,\alpha_N) ~=~\mbox{const.}$, which
has to satisfy the partial differential equation \newline
${\vec \beta}\cdot {\vec \nabla}\,\Phi ~=~\sum_{i=0}^{N}
\,\beta_{i}\,(\partial\Phi/\partial g_{i})~=~0$,
where $\beta_i$ is the $\beta$-function of $\alpha_i$.
In general,
 there exist $N$ independent solutions of them,
and they are equivalent to the solutions of
the so-called reduction equations \cite{zimmermann1},
\be
\beta \,\frac{d \alpha_{i}}{d \alpha} &=&\beta_{i}~,~i=1,\cdots,N~,
\ee
where $\alpha\equiv \alpha_0$ and
$\beta\equiv\beta_0 $.
Since maximally $N$ independent
RG invariant constraints
in the $(N+1)$-dimensional space of couplings
can be imposed by $\Phi_i$, one could in principle
express all the couplings in terms of
a single coupling $\alpha$  \cite{zimmermann1}.
This is the basic observation to understand reduction of couplings.

We assume that the evolution equations of $\alpha$'s take the form
\be
\frac{d \alpha}{dt} &=&-b^{(1)}\,\alpha^2+\cdots~,\nn\\
\frac{d\alpha_i}{dt} &=&-b^{(1)}_{i}\,\alpha_{i}\alpha+
\sum_{j,k}b^{(1)}_{i,jk}\,\,\alpha_j\alpha_k+\cdots~,
\ee
in perturbation theory, and then we derive from (2)
\be
\alpha \frac{d \tilde{\alpha}_{i}}{d\alpha} &=&
(\,-1+\frac{b^{(1)}_{i}}{b^{(1)}}\,)\, \tilde{\alpha}_{i}
-\sum_{j,k}\,\frac{b^{(1)}_{i,jk}}{b^{(1)}}
\,\tilde{\alpha}_{j}\, \tilde{\alpha}_{k}+\sum_{r=2}\,
(\frac{\alpha}{\pi})^{r-1}\,\tilde{b}^{(r)}_{i}(\tilde{\alpha})~,
\ee
where $\tilde{\alpha}_i \equiv \alpha_i/\alpha$ and
$\tilde{b}^{(r)}_{i}(\tilde{\alpha})$ with $r=2,\cdots$
are power series of $\tilde{\alpha}_{i}$ and can be computed
from the $r$-th loop $\beta$-functions ($i=1,\cdots,N$).
We next solve  the algebraic equations
\be
(\,-1+\frac{b^{(1)}_{i}}{b^{(1)}}\,)\, \rho_{i}
-\sum_{j,k}\frac{b^{(1)}_{i,jk}}{b^{(1)}}
\,\rho_{j}\, \rho_{k}&=&0~,
\ee
and assuming that their solutions $\rho_{i}$'s have the form
\be
\rho_{i}&=&0~\mbox{for}~ i=1,\cdots,N'~;~
\rho_{j} ~>0 ~\mbox{for}~j=N'+1,\cdots,N~,
\ee
we regard $\tilde{\alpha}_{i}$ with $i \leq N'$
 as small
perturbations.  The
undisturbed system is defined by setting
all $\tilde{\alpha}_{i}$  with $i \leq N'$ equal to zero.
It is possible
\cite{zimmermann1} to verify at the one-loop level
the existence of
the unique power series solutions
\be
\tilde{\alpha}_{j}&=&\rho_{j}+\sum_{r=2}\rho^{(r)}_{j}\,
(\frac{\alpha}{\pi})^{r-1}~,~j=N'+1,\cdots,N~
\ee
to the reduction equations (2) to all orders
in  the undisturbed system.
These are the RG invariant relations among couplings that keep
formally perturbative renormalizability of the undisturbed system.
We emphasize that the more
vanishing $\rho_i$'s a solution contains, the less is its
predictive power in general. We, therefore, search for predictive
solutions in a systematic fashion.

 The small
 perturbations caused by nonvanishing $\tilde{\alpha}_{i}$'s
 with $i \leq N'$ defined above
 enter in such a way that the reduced
couplings, i.e., $\tilde{\alpha}_{i}$  with $i > N'$,
become functions not only of
$\alpha$ but also of $\tilde{\alpha}_{i}$
 with $i \leq N'$.
It turned out \cite{kubo4,kubo2} that, to investigate such partially
reduced systems, it is most convenient to work with a set of
partial differential equations
\be
\{~~\tilde{\beta}\,\frac{\partial}{\partial\alpha}
+\sum_{k=1}^{N'}\,
\tilde{\beta_{k}}\,\frac{\partial}{\partial\tilde{\alpha}_{k}}~~\}~
\tilde{\alpha}_{j}(\alpha,\tilde{\alpha})
&=&\tilde{\beta}_{j}(\alpha,\tilde{\alpha})~,
{}~j=N'+1,\cdots, N,~,\\
\tilde{\beta}~\equiv~\frac{\beta}{\alpha}~~,~~
\tilde{\beta}_{i} &=&\frac{\beta_{i}}{\alpha^2}
-\frac{\beta}{\alpha^{2}}~\tilde{\alpha}_{i}~,~i=1,\cdots,N~.\nn
\ee
The partial differential equations  (8) are equivalent
to the reduction equations (2), and we
look for their solutions in the form
\be
\tilde{\alpha}_{j}&=&\rho_{j}+
\sum_{r=2}\,(\frac{\alpha}{\pi})^{r-1}\,f^{(r)}_{j}
(\tilde{\alpha}_{i})~,
\ee
where $ f^{(r)}_{j}$ are supposed to be
power series of
$\tilde{\alpha}_{i}~,~i=1,\cdots,N'$.
This particular type of solutions
can be motivated by requiring that in the limit of vanishing
perturbations we obtain the undisturbed
solutions (7), i.e.,
$ f_{j}^{(r)}(0)=\rho_{j}^{(r)}~\mbox{for}~r \geq 2$.
Again it is possible to obtain  the sufficient conditions for
the uniqueness of $ f^{(r)}_{j}$ in terms of the lowest order
coefficients.

So this is the machinery to build gauge-Yukawa unified models.
In the next section, we consider an explicit example.

\section{An example: The minimal susy $SU(5)$ GUT}

To illustrate our method of GYU, we consider \cite{kubo4} the minimal
$N=1$ supersymmetric
GUT based on the group $SU(5)$ \cite{sakai1}.
As well-known, three generations of quarks and leptons
are accommodated by
three chiral supermultiplets in
$\Psi^{I}({\bf 10})$ and $\Phi^{I}(\overline{\bf 5})$,
where $I$ runs over the three generations.
One uses a $\Sigma({\bf 24})$ to break $SU(5)$ down to $SU(3)_{\rm C}
\times SU(2)_{\rm L} \times U(1)_{\rm Y}$,  and
$H({\bf 5})$ and $\overline{H}({\overline{\bf 5}})$
 to describe the
two Higgs supermultiplets appropriate for
electroweak symmetry breaking.
The superpotential of the model is given by
\be
W &=& \frac{g_{t}}{4}\,
\epsilon^{\alpha\beta\gamma\delta\tau}\,
\Psi^{3}_{\alpha\beta}\Psi^{3}_{\gamma\delta}H_{\tau}+
g_b\,\overline{H}^{\alpha}
\Psi^{3}_{\alpha\beta}\Phi^{3 \beta}+
+\frac{g_{\lambda}}{3}\,\Sigma_{\alpha}^{\beta}
\Sigma_{\beta}^{\gamma}\Sigma_{\gamma}^{\alpha}+
g_{f}\,\overline{H}^{\alpha}\Sigma_{\alpha}^{\beta} H_{\beta}\nn\\
& &+ m_{1}\, \Sigma_{\alpha}^{\gamma}\Sigma_{\gamma}^{\alpha}+
+m_{2}\,\overline{H}^{\alpha} H_{\alpha}~,
\ee
where $\alpha,\beta,\cdots$ are the $SU(5)$
indices, and we have suppressed the Yukawa couplings of the
first two generations.
The one-loop $\beta$-functions of the  couplings in $W$  are found to
be
\be
\beta^{(1)} &=& -\frac{3}{16\pi^2}\,g^3~,\nn\\
\beta^{(1)}_{t} &=& \frac{1}{16\pi^2}\,
[\,-\frac{96}{5}\,g^2+9\,g_{t}^{2}+\frac{24}{5}\,g_{f}^{2}+
4\,g_{b}^{2}\,]\,g_{t}~,\nn\\
\beta^{(1)}_{b} &=& \frac{1}{16\pi^2}\,
[\,-\frac{84}{5}\,g^2+3\,g_{t}^{2}+\frac{24}{5}\,g_{f}^{2}+
10\,g_{b}^{2}\,]\,g_{b}~,\\
\beta^{(1)}_{\lambda} &=& \frac{1}{16\pi^2}\,
[\,-30\,g^2+\frac{63}{5}\,g_{\lambda}^2+3\,g_{f}^{2}
\,]\,g_{\lambda}~,\nn\\
\beta^{(1)}_{f} &=& \frac{1}{16\pi^2}\,
[\,-\frac{98}{5}\,g^2+3\,g_{t}^{2}
+4\,g_{b}^{2}
+\frac{53}{5}\,g_{f}^{2}+\frac{21}{5}\,g_{\lambda}^{2}
\,]\,g_{f}~.\nn
\ee
According to the notation introduced in the previous section, we
define \newline
  $\tilde{\alpha}_{i}~\equiv~
\alpha_{i}/\alpha~,~
\alpha_{i} ~=~|g_{i}|^{2}/4\pi~,~i=t,b,\lambda,f$.

There may exist in principle $2^4=16$
non-degenerate  solutions to the algebraic equations
(5), corresponding  to vanishing $\rho$'s
as well as nonvanishing ones as given in (6).
Here we require the
solutions to be most predictive ($\rho_t,\rho_b \neq 0$)
and to describe an asymptotically free
system.  One finds \cite{kubo4} that there exist exactly  two
solutions that satisfy these requirements:
\be
1 &:& \rho_{t}=\frac{2533}{2605}~,~
\rho_{b}=\frac{1491}{2605}~,~\rho_{\lambda}=0~,~
\rho_{f}=\frac{560}{521}~,\\
2 &:& \rho_{t}=\frac{89}{65}~,~
\rho_{b}=\frac{63}{65}~,~\rho_{\lambda}=0~,~
\rho_{f}=0~.\nn
\ee
On can also show that for both cases the corresponding power series
solutions of the form (7) uniquely exist.

Further,
according to the previous section, both solutions
 give the possibility to obtain  partial reductions,
where $\tilde{\alpha}_{\lambda}$ has to be
regarded as the small perturbation
in the case of solution 1, and $\tilde{\alpha}_{\lambda}$
and $\tilde{\alpha}_{f}$ are those for solution 2.
Corrections in  lower orders are found to be
\be
\tilde{\alpha}_{j} &=& \rho_{j}+
f^{(r_{\lambda}=1)}_{j}\,
\tilde{\alpha}_{\lambda}+f^{(r_{\lambda}=2)}_{j}\,
\tilde{\alpha}_{\lambda}^2+\cdots~~,~j=t,b,f~,
\ee
for  solution $1$, where
\be
f^{(r_{\lambda}=1)}_{t,b,f} &\simeq &  0.018~,~0.012~,~-0.131~,~
f^{(r_{\lambda}=2)}_{t,b,f} \simeq  0.005~,~0.004~,~-0.021~,\nn
\ee
and for solution $2$,
\be
\tilde{\alpha}_{j} &=& \rho_{j}+f^{(r_{f}=1)}_{j}\,
\tilde{\alpha}_{f}+f^{(r_{\lambda}=1)}_{j}\,
\tilde{\alpha}_{\lambda}+f^{(r_{f}=1,r_{\lambda}=1)}_{j}\,
\tilde{\alpha}_{f}\,\tilde{\alpha}_{\lambda}\nn\\
& &+f^{(r_{f}=2)}_{j}\,
\tilde{\alpha}_{f}^2+f^{(r_{\lambda}=2)}_{j}\,
\tilde{\alpha}_{\lambda}^{2}\cdots~~,~j=t,b~,
\ee
where
\be
f^{(r_{\lambda}=1)}_{j}&=&
f^{(r_{\lambda}=2)}_{j}~=~0~,~
f^{(r_{f}=1)}_{t,b}
\simeq  -0.258~,~-0.213~,\nn\\
f^{(r_{f}=2)}_{t,b}
&\simeq &-0.055~,~ -0.050~,~
f^{(r_{f} =1,r_{\lambda}=1)}_{t,b}
\simeq  -0.021~,~-0.018 ~.\nn
\ee
A detailed analysis \cite{kubo4} shows that to keep asymptotic
freedom  in the case of
solution 2,  the $ \tilde{\alpha}_{\lambda}$ is allowed
to vary  from $0$ to $15/7$  while the
$\tilde{\alpha}_{f} $ may vary
from $0$ to a maximal value which depends on  $
\tilde{\alpha}_{\lambda}$ (in the one-loop
approximation). One furthermore finds that
solution 1 is the boundary of solution 2 so that both solutions
belong to the same RG invariant, asymptotically free
 surface. This is
shown in Fig. 1.
  \begin{figure}
           \epsfxsize= 9 cm   
           \centerline{\epsffile{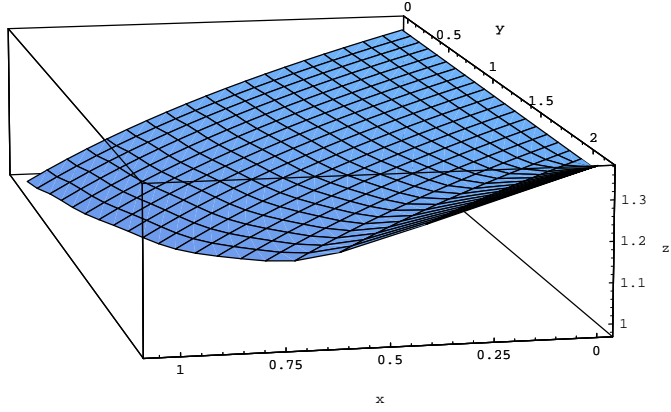}}
        \caption{Asymptotically free surface, where the $x,~y$ and
$z$ axes correspond to $\tilde{\alpha}_{f},~
\tilde{\alpha}_{\lambda}$ and $\tilde{\alpha}_{t}$, respectively.}
        \label{fig:1}
        \end{figure}

Eq. (14) defines
GYU boundary conditions holding
at $M_{\rm GUT}$.
Note that they remain unaffected by
soft supersymmetry breaking terms,
because the $\beta$-functions are
not altered by these terms.
 To predict observable parameters from GYU,
we apply the well-known
RG technique.
We  assume that below $M_{\rm GUT}$ the evolution of couplings is
 governed by the MSSM and that there exists a unique threshold
$M_{\rm SUSY}$ for all superpartners of the MSSM so that
below $M_{\rm SUSY}$ the SM is the correct effective
theory.

We emphasize that with a GYU boundary condition
alone the value of $\tan\beta$ can not be determined.
Usually, it is determined in the Higgs sector, which however
strongly depends on the supersymmetry braking terms.
Here we avoid this by using the tau mass $M_{\tau}$ as an input.
As the input data we
use $M_{\tau}=1.777~\mbox{GeV}~,~
M_{Z}=91.188~\mbox{GeV}~,~
\alpha_{\rm em}^{-1}(M_{Z}) = 127.9
+(8/9\pi)\,\log (M_t/M_Z) ~,~
\sin^{2} \theta_{\rm W}(M_{Z})=0.2319
-3.03\times 10^{-5}T-8.4\times 10^{-8}T^2$, where
$T =  M_t /[\mbox{GeV}] -165$.
As we can see from (14),
the $\tilde{\alpha}_{\lambda}$-dependence
of $\tilde{\alpha}_{t}$
and $\tilde{\alpha}_{b}$ are rather
weak, and so we present in Table 1 the
predictions for three different values of
$\tilde{\alpha}_{f}$ with $\tilde{\alpha}_{\lambda}$ fixed at zero
only ($M_{\rm SUSY}=500$ GeV):\footnote{Small
$\tilde{\alpha}_{\lambda}$
is preferable because of the nucleon decay
constraint as we will see later.}

\begin{center}
{\bf Fig. 1}
\end{center}

\vspace{0.2cm}
\begin{tabular}{|c|c|c|c|c|c|c|c|}
\hline
$\tilde{\alpha}_{f}$
 & $\tilde{\alpha}_{t}$ & $\tilde{\alpha}_{b}$&
$\alpha_{3}(M_Z)$ &
$\tan \beta$  &  $M_{\rm GUT}$ [GeV]
 & $m_b (M_{b}) $ [GeV]& $M_{t}$ [GeV]
\\ \hline
$0$ & $1.369$ & $0.969$ & $0.1217$  &$52.6$  &
$1.76\times 10^{16}$  & $4.59$  & $182.6$ \\ \hline
 $0.6$ & $1.187$& $0.816$ & $0.1216$  & $51.1$  & $1.75\times 10^{16}$
  & $4.64$  & $181.0$  \\ \hline
$1.075$ & $0.972$& $0.572$ & $0.1216$  & $47.9$   & $1.73\times 10^{16}$
  &  $4.72$ & $179.0$   \\ \hline
\end{tabular}

\vspace{0.2cm}
\noindent
$M_{t,b}$ are the pole masses while $m_b (M_b)$
is the running bottom quark
mass at its pole mass.
The values for $m_b (M_b)$ may suffer from
 a relatively large
correction coming from the superpartner contribution
which is not included above.
Because of the
infrared behavior of the Yukawa couplings \cite{hill1},  the
value of $M_t$ may be  insensitive  against the change
of $\tilde{\alpha}_{t}$. In Fig. 2 we plot
$M_t$ against $ \tilde{\alpha}_{t}$ with
$\tilde{\alpha}_{b}=0.642$ and
$M_{\rm SUSY}=500$ GeV
fixed,\footnote{A similar analysis has been done
by Bando {\em et al}. in Ref. \cite{bando} on the one-loop level,
but not to study GYU physics.} where the reduction solution
corresponds to $\tilde{\alpha}_{t}=1.0$.
 \begin{figure}
           \epsfxsize= 9 cm   
           \centerline{\epsffile{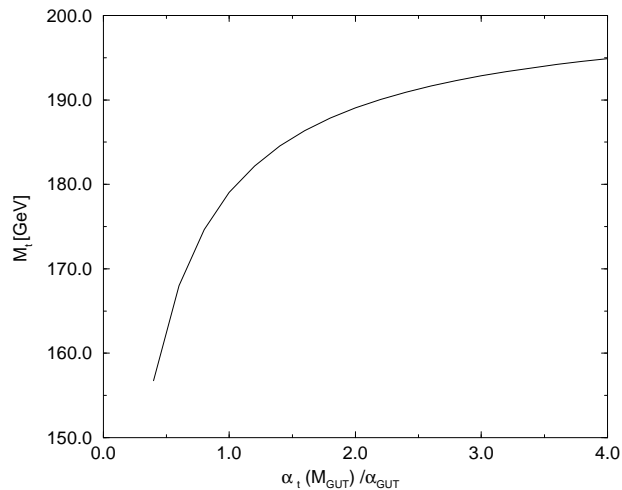}}
        \caption{$M_t$ versus $ \tilde{\alpha}_{t}$
 with fixed $ \tilde{\alpha}_{b}$ and $M_{\rm SUSY}$.
The reduction solution corresponds to $\tilde{\alpha}_{t}=1 $.}
        \label{fig:2}
        \end{figure}
{}From Fig. 2 we see that
with increasing experimental accuracy of $M_t$ it may
become possible to test various GYU models.
Detailed studies on this problem will be
published
elsewhere \cite{kubo6}.

\vspace{0.2cm}
Finally we would like to turn our discussion to proton decay.
Since the couplings in the minimal model with
GYU are strongly constrained, the parameter space for proton decay
is also constrained.  To see this, we recall that if one includes
the threshold effects of superheavy particles \cite{threshold},
the GUT scale $M_{\rm GUT}$ at which $\alpha_1$ and $\alpha_2$
meet is related to the mass of the superheavy
$SU(3)_C $-triplet  Higgs supermultiplets contained
in $H$ and $\overline{H}  $ by
\be
M_H &=&[\tilde{\alpha}_{f}]^{15/28}\,
[\tilde{\alpha}_{\lambda}]^{-5/28}\,M_{\rm GUT}~.
\ee
This mass $M_H$ controls the nucleon decay which
is mediated by dimension five operators \cite{d5}, and
non-observation of the nucleon decay requires
$M_H \gsim 10^{17}$ GeV for $\tan\beta\sim 50$ \cite{hisano1}.
Since $M_{\rm GUT} \sim 1.7\times 10^{16}$ GeV and
$\tilde{\alpha}_{f} \lsim 1.1 $ as
one can see from eq. (14) and Table 1, the value of
$\tilde{\alpha}_{\lambda}$ has to be less than
$\sim 4.4 \times 10^{-5}$. Therefore, the reduction solutions that
are consistent with the nucleon decay constraint are very close to
 solution 1, so to the boundary of the asymptotically
surface shown in Fig. 2.

\section{Dynamical Unification of Couplings}

As we have seen, we can construct gauge-Yukawa unified models
by applying the
principle of reduction of couplings. Though there
are certain successes of these models, the reduction principle is
associated with no  intuitive, physical meaning.
Dynamical Unification of Couplings \cite{kubo3}, which
we are going to explain, could give a reduction of couplings
a simple, theoretical meaning. There exists already an example
of DUC: Triviality of gauged Higgs-Yukawa systems is
widely expected, unless they are completely asymptotically free.
It was found\cite{kubo4} that
 by imposing  a certain
relation among the gauge, Higgs  and Yukawa couplings
which are consistent with perturbative renormalizability,
it is possible to make
the $SU(3)$-gauged Higgs-Yukawa system
completely  asymptotically free
and hence nontrivial.\footnote{It has been found by
Harada {\em et al}. in Ref. \cite{harada}
that asymptotic freedom of
gauged Higgs-Yukawa systems is closely related to the
nonperturbative existence of gauged
Nambu-Jona-Lasino models. The models have been considered in
the ladder approximation by Kondo
{\em et al}. in Ref. \cite{kondo} before, who
also have observed a DUC in the models. }
This RG invariant
relation among couplings
is a consequence of the reduction of couplings.
A DUC appears if the couplings in a theory are forced
in a dynamically consistent fashion to be related
with each other in order
for the theory to remain  well-defined in the ultraviolet limit.

\begin{figure}
           \epsfxsize= 8 cm   
           \centerline{\epsffile{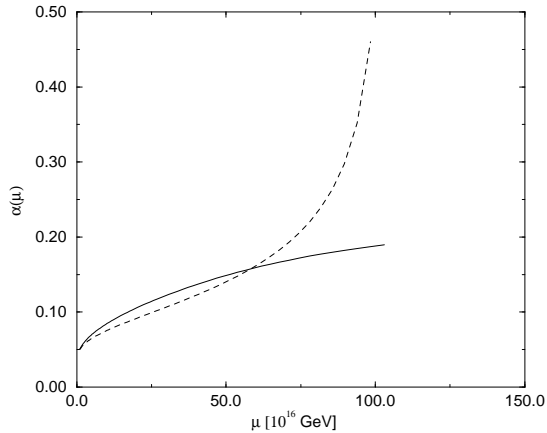}}
        \caption{The evolution of $\alpha$ above $M_{\rm GUT}$;
 in the $\overline{\rm MS}$ scheme (dashed) and  in OPT
(solid line).}
        \label{fig:3}
        \end{figure}

However, most  grand unified theories  become
asymptotically nonfree,
if  one attempts to obtain a desired symmetry breaking pattern and
a realistic
fermion mass matrix by introducing
more Higgs fields.
The common wisdom is that
such theories develop a Landau pole at a high energy
scale, a fact which inevitably suggests that the theory is trivial,
unless some new physics is entering before the couplings blow up.
There exist, however,
arguments \cite{stevenson2,kubo3} based on
optimized perturbation theory (OPT) \cite{stevenson1},
indicating that non-abelian gauge theories can have a
nontrivial fixed
 point.\footnote{The existence of a nontrivial
ultraviolet fixed point in Yang-Mills theories nicely
fit with the idea of walking
technicolor gauge coupling  \cite{technicolor}}.
If this is the
case, the idea of DUC could be applied to asymptotically nonfree
theories, too.

Unification of the gauge couplings in
asymptotically nonfree extensions of the
SM were previously considered \cite{maiani}.
In contrast to the present idea, it was assumed
that the gauge couplings asymptotically diverge
 so that  if one requires
the couplings to become strong
simultaneously at a certain energy scale,
one can predict their low energy values.
The difference of two approaches may be illustrated  in Fig. 3, which
shows the three-loop evolution of the
 gauge coupling $\alpha$ above $M_{\rm GUT}$ scale
in the $\overline{\rm MS}$ scheme and in
optimized perturbation theory in the $SO(10)$-gauge theory with
$30$ Dirac fermions in the fundamental representation.
The existence of a nontrivial fixed point found in OPT
and also the possibility of DUC
have to be independently verified
in different approaches, of course.


\newpage
\begin{thebibliography}{99}
\bi{gut1} J.~C.~Pati and A.~Salam, Phys.~Rev.~Lett.\ {\bf 31}
(1973) 661; H.~Georgi and S.~L.~Glashow, Phys.~Rev.~Lett.\
{\bf 32} (1974) 438.
\bi{kubo1} J.~Kubo, K.~Sibold and W.~Zimmermann,
Nucl.~Phys.\ {\bf B259} (1985) 331.
\bi{mondragon1}D.~Kapetanakis, M.~Mondrag{\' o}n and
G.~Zoupanos, Z.~Phys.\ {\bf C60} (1993) 181;
 M.~Mondrag{\' o}n and G.~Zoupanos,
Nucl.~Phys.\ {\bf B} (Proc. Suppl)
 37 {\bf C} (1995) 98.
\bi{kubo2} J.~Kubo, M.~Mondrag{\' o}n and
G.~Zoupanos, Nucl.~Phys.\ {\bf B424} (1994) 291.
\bi{finite1}
A.~J.~Parkes and P.~C.~West, Phys.~Lett.\ {\bf 138B}
(1984) 99; Nucl.~Phys.\ {\bf B256} (1985) 340;
D.~R.~T.~Jones and A.~J.~Parkes, Phys.~Lett.\ {\bf B160}
(1985) 267;
D.~R.~T. Jones and L.~Mezinescu, Phys.~Lett.\ {\bf B136}
(1984) 242; {B138}  (1984) 293;
A.~J.~Parkes, Phys.~Lett.\ {\bf B156} (1985) 73;
S.~Hamidi, J.~Patera and J.~H.~Schwarz, Phys.~Lett.\ {\bf B141}
(1984) 349;
D.~R.~T.~Jones and S.~Raby,
Phys.~Lett.\ {\bf B143} (1984) 137;
S.~Hamidi and J.~H.~Schwarz,
Phys.~Lett.\ {\bf B147} (1984) 301;
J.~E.~Bj{\" o}rkman, D.~R.~T.~Jones and S.~Raby,
Nucl.~Phys.\ {\bf B259} (1985) 503;
J. Le{\'  o}n et al, Phys.~Lett.\ {\bf B156} (1985) 66;
X.~D.~Jiang and X.~J.~Zhou,
Phys.~Lett.\ {\bf B197} (1987) 156; {\bf B216} (1989) 160;
I.~Jack and D.~R.~T.~Jones, Phys.~Lett.\ {\bf B333} (1994) 372.
\bi{finite2}
A.~V.~Ermushev, D.~I.~Kazakov and O.~V.~Tarasov,
Nucl.~Phys.\ {\bf B281} (1987) 72;
D.~I.~Kazakov, Mod.~Phys.~Lett.\ {\bf A2} (1987) 663;
 Phys.~Lett.\ {\bf B179} (1986) 352;
 C.~Lucchesi, O.~Piguet and K.~Sibold,
Helv.~Phys.~Acta. {\bf 61} (1988) 321.
\bi{zimmermann1}W.~Zimmermann, Commun.~Math.~Phys.\ {\bf 97}
(1985) 211;
R.~Oehme and W.~Zimmermann, Commun.~Math.~Phys.\ {\bf 97}
(1985) 569.
\bi{gut2}H.~Georgi, H.~Quinn, S.~Weinberg, Phys.~Rev.~Lett.\
{\bf 33} (1974) 451.
\bi{kubo3} J.~Kubo, {\em Nontrivial Asymptotically
Nonfree Gauge Theories and Dynamical Unification
of Couplings}, to appear in Phys.~Rev.\ {\bf D}.
\bi{sakai1} S.~Dimopoulos and H.~Georgi, Nucl.~Phys.\ {\bf B193}
(1981) 150;
 N.~Sakai, Z.~Phys.\ {\bf C11} (1981) 153.
\bi{hill1}C.~T.~Hill, Phys.~Rev.\ {\bf D24} (1981) 691;
C.~T.~Hill, C.~N.~Leung and S.~Rao, Nucl.~Phys.\ {\bf B262}
(1985) 517;
W.~A.~Bardeen, M.~Carena, S.~Pokorski and C.~E.~M.~Wagner,
Phys.~Lett.\ {\bf B320} (1994) 110.
\bi{bando}M.~Bando, T.~Kugo, N.~Maekawa and H.~Nakano,
Mod.~Phys.~Lett.\ {\bf A7} (1992) 3379.
\bi{kubo6}J.~Kubo, M.~Mondrag{\' o}n,
M.~Olechowski and G.~Zoupanos, to appear.
\bi{threshold}J.~Hisano, H.~Murayama and T.~Yanagida,
Phys.~Rev.~Lett. {\bf 69} (1993) 1992;
J.~Ellis,S.~Kelley and D.~V.~Nanopoulos,
Nucl.~Phys.\ {\bf B373} (1992) 55;
Y.~Yamada, Z.~Phys.\ {\bf C60} (1993) 83.
\bi{d5}N.~Sakai and T.~Yanagida, Nucl.~Phy.\ {\bf B197}
(1982) 533; S.~Weinberg, Phys.~Rev.\ {\bf D26} (1982) 287.
\bi{hisano1}J.~Hisano, H.~Murayama and T.~Yanagida,
Nucl.~Phys.\ {\bf B402} (1993) 46.
\bi{kubo4} J.~Kubo, K.~Sibold and W.~Zimmermann,
Phys.~Lett.\ {\bf 220B} (1989) 185; J.~Kubo,
Phys.~Lett.\ {\bf 262B} (1991) 472.
\bi{harada}M.~Harada {\em et al}., Prog.~Theor.~Phys.\ {\bf 92},
(1994) 1161.
\bi{kondo}K.~-I.~Kondo, M.~Tanabashi and K.~Yamawaki,
Prog.~Theor.~Phys.\ {\bf 89} (1993)1249;
K.~-I.~Kondo {\em et al}., Prog.~Theor.~Phys.\ {\bf 91}
(1994) 541.
\bi{stevenson2} A.~C.~Mattingly and P.~M.~Stevenson,
Phys.~Rev.~Lett.\ {\bf 69} (1992) 1320;
Phys.~Rev.\ D {\bf D49} (1994) 47.
\bi{stevenson1}P.~M.~Stevenson,
Phys.~Rev.\ {\bf D23} (1981) 2916; J.~Kubo, S.~Sakakibara
 and P.~M.~Stevenson,  Phys.~Rev.\ {\bf D29}, 1682  (1984).
\bibitem{technicolor}
B.~Holdom, Phys.~Lett.\  {\bf 150B} (1985) 301;
K.~Yamawaki, M.~Bando and K.~Matumoto,
Phys.~Rev.~Lett.\ {\bf 56}  (1986) 1335;
T.~Akiba, and T.~Yanagida,
Phys.~Lett.\  {\bf 169B} (1986) 432;
T.~Appelquist, D.~Karabali and L.~C.~R.~Wijewardhana,
Phys.~Rev.~Lett.\ {\bf 57} (1986) 957.
\bibitem{maiani}
L.~Maiani, G.~Parisi and R.~Petronzio,
Nucl.~Phys.\ {\bf B136}
(1978) 115; N.~Cabibo and G.~R.~Farrar, Phys.~Lett.\
{\bf 110B} (1982) 107;
 L.~Maiani and R.~Petronzio, Phys.~Lett.\ {\bf B176} (1986) 12;
G.~Grunberg, Phys.~Rev.~Lett.\ {\bf 58} (1987) 1180;
S.~Theisen, N.~D.~Tracas
and G.~Zoupanos, Z.~Phys. {\bf C37} (1988) 597;
C.~Alacoque, C.~Deom and J.~Pestieau, Phys.~Lett.\ {\bf B228} (1989)
370; D.~Kapetanakis, S.~Theisen and G.~Zoupanos,
Phys.~Lett.\ {\bf B229} (1989) 248;
P.~Q~Hung, Phys.~Rev.\ {\bf D38} (1988) 377;
J.~P.~Derendinger, R.~Kaiser
and M.~Roncadelli, Phys.~Lett.\ {\bf B220} (1989) 164;
G.~Leontaris, C.~E.~Vayonakis and N.~D.~Tracas,
Mod.~Phys.~Lett.\ {\bf A4} (1989) 2429;
T.~Mori, H.~Murayama and T.~Yanagida, Phys.~Rev.\ {\bf D48}
(1993) 2995;
 B.~Brahmachari, U.~Sarkar and K.~Sridhar,
Mod.~Phys.~Lett.\ {\bf A8} (1993) 3349.

\end{thebibliography}
\end{document}